\documentclass[12pt]{iopart}

\newcommand {\be}{\begin{equation}} % start equation
\newcommand{\ee}{\end{equation}}    % end equation
\def\ddt{\frac{\partial}{\partial t}}

\def\dds1{\frac{\partial}{\partial s_1}}

\def\d{d\kern-0.8 ex\vrule height 1.3 ex depth-1.24 ex width 0.7 ex
\kern 0.15 ex}
\def\D{D\kern-1.7 ex\vrule height .87 ex depth-0.8 ex width 0.7 ex
\kern 0.95 ex}

\begin{document}

\title[Nonlinear compressive  magnetoacoustic Alfv\'{e}nic waves]{Some unexplored features of the nonlinear compressive  magnetoacoustic Alfv\'{e}nic waves}

\author{J. Vranjes$^1$, B. P. Pandey$^{2}$}

\address{$^{1}$Institute of Physics Belgrade, Pregrevica 118, 11080 Zemun, Serbia.\\

 $^{2}$Department of Physics \& Astronomy \& Research Centre for Astronomy, Astrophysics \& Astrophotonics, Macquarie University, Sydney, NSW
2109, Australia.\\
}

\ead{jvranjes@yahoo.com; drbp.pandey@gmail.com}
\begin{abstract}
The theory of nonlinear magnetoacoustic wave in the past has strictly been focused on purely compressive features of the mode. We show that a complete set of nonlinear equations necessarily includes both compressional and shear components of the magnetic field. These two turn out to be described by exactly the same nonlinear equations, which make the use of such a complete full set of equations far less complicated than expected. Present results should considerably enrich the theory of these waves by opening up new frontiers of investigation and providing some completely new types of nonlinear solutions.
\end{abstract}

%Uncomment for PACS numbers title message
% Keywords required only for MST, PB, PMB, PM, JOA, JOB?
%\vspace{2pc}
%\noindent{\it Keywords}: Article preparation, IOP journals
% Uncomment for Submitted to journal title message
%\submitto{\JPA}
% Comment out if separate title page not required
\maketitle

\section{Introduction}

There have been many studies of the nonlinear hydro-magnetic waves in the past half a century of plasma science. This is not surprising in view of the fact that such waves are expected in various environments in the laboratory, space and astrophysical plasmas. The theory of such waves  can be traced as far back as to  1958 \cite{aa}, and it can be seen in many studies ever since \cite{w}-\cite{shi}.

In the papers \cite{aa}-\cite{shi} electromagnetic perturbations propagating perpendicular to the magnetic field are assumed  to be purely compressible (i.e., having the perturbed magnetic field component only along the ambient magnetic field vector). Within the linear theory, such  purely  compressive magnetic perturbations indeed follow from the geometry of the mode, and this  holds without any assumption. However, within the nonlinear theory an additional perpendicular shear component of the magnetic field appears naturally.   This perpendicular component  turns out to be described by an equation which is {\em exactly the same} as the equation for the compressional component, see later in the text. Nevertheless, the usual nonlinear  theory \cite{aa}-\cite{shi} dealing with purely compressive phenomena  is formally correct, such compressive nonlinear solutions are mathematically allowed and physically possible. Physical phenomena obtained within such a model are numerous and this is partly seen from the huge number of studies in the past fifty years, and  after so many decades the  mode is still in the focus of researchers  as it was in its early days, c.f. recent Refs.~\cite{m2}-\cite{shi}.  The unexplored  physics of the nonlinear mode, that should follow from  the additional shear component, is expected to be important.

 In this work a closed  set of nonlinear equations for the generalized magnetoacoustic mode will be derived showing that both compressible and shear components of the magnetic field are described by an exactly the same nonlinear equation. One simple solution of such a complete set of equations will be presented in order to show that  such a generalized theory can be used in a manner very similar to the usual studies where only the compressional part of the magnetic field is taken into account.

\section{Model and equations}

 The model which we describe here uses an arbitrary  background magnetic field   in the $z$-direction $\vec B_0= B_0 \vec e_z$. Further, we shall use two-fluid description for the electrons and ions that include, respectively, the momentum equations for the two species $j=e, i$, and the Faraday and Amp\`{e}re laws:
 \be
 \frac{D_e\vec v_e}{Dt}=- \frac{e}{ m_e} \vec E -\nabla p_e
   - \frac{e}{m_e} \vec v_e\times \vec B  -  \nu_{ei} (\vec v_e- \vec v_i),
   \label{ee1}
     \ee
  \be
 \frac{D_i\vec v_i}{Dt}= \frac{e}{ m_i} \vec E -\nabla p_i
   + \frac{e}{ m_i } \vec v_i\times \vec B  -  \nu_{ie} (\vec v_i- \vec v_e),
   \label{ee1a}
    \ee
   \be
 \nabla\times \vec E=-\frac{\partial \vec B}{\partial t}, \label{ee3}
  \ee
 \be
 \nabla\times \vec B=\mu_0 \vec j= \mu_0 e \left(n_i \vec v_i-n_e \vec v_e\right). \label{ee4}
 \ee
In Eqs.~(\ref{ee1})-(\ref{ee4}) singly charged ions (i.e., protons) are assumed. We shall use  the quasi-neutrality condition:
  \be
 n_e=n_i=n, \label{e5}
 \ee
 and the continuity equation for ions
 \be
 \frac{\partial n}{\partial t} + \nabla\cdot\left(n \vec v_i\right)=0.\label{con}
 \ee
 We shall discuss perturbations propagating through a homogeneous  plasma in the $x$-direction, perpendicular to the ambient  magnetic field so that $D_j/Dt\equiv\partial/\partial t+ v_{jx} \partial/\partial x$.

%\section{Small linear perturbations}
To point out some basic properties of the mode, in this part  we omit collisions, and assume  inertia-less electrons and isothermal and quasi-neutral perturbations $p_{j1}=\kappa T_j n_1$.  Within these assumptions, from linearized Eq.~(\ref{ee1}), one may calculate the electric field $\vec E_1$
\be
\vec E_1=-\frac{\kappa T_e}{e n_0} \nabla n_1 - \vec v_{e1}\times \vec B_0, \label{e6}
\ee
and plug it in the momentum equation for ions, which, with the help of linearized Amp\`{e}re law [Eq.~\ref{ee4}] consequently becomes
\be
m_in_0 \frac{\partial \vec v_{i1}}{\partial t}=\frac{1}{\mu_0} \left(\nabla\times \vec B_1\right) \times \vec B_0- \kappa\left(T_i+ T_e\right)\nabla n_1. \label{e7}
\ee
Taking the curl of Eq.~(\ref{e6}) and expressing $\vec v_{e1}$ from the linearized  Eq.~(\ref{ee4}) with the help of Eq.~(\ref{ee3})
one obtains the following induction equation
\be
\frac{\partial \vec B_1}{\partial t}= \nabla\times \left(\vec v_{i1}\times \vec B_0\right)-\nabla \times \left[\frac{1}{\mu_0 e n_0} \left(\nabla\times \vec B_1\right) \times \vec B_0\right]. \label{e8}
\ee
The linear set of equations is closed with the linearized  version of Eq.~(\ref{con}). For perturbations propagating in the $x$-direction $\nabla\equiv \vec e_x \partial/\partial x$, the first   term on the right-hand side in (\ref{e8}) yields $- B_0 \vec e_z \partial v_{ix1}/\partial x$, while the second term vanishes.
Hence, Eq.~(\ref{e8}) yields only the $z$-component of $\vec B_1$. Similarly, the first term on the right-hand side in Eq.~(\ref{e7}) also contains only the $z$-component of $\vec B_1$ because  $ \left(\nabla\times \vec B_1\right) \times \vec B_0=-\vec e_x B_0 \partial B_{z1}/\partial x$.

Within the linear regime, equations  consequently yield the hydrodynamically longitudinal ($\vec v_{i1}\equiv v_{ix1} \vec e_x$) and electro-dynamically transverse ($\vec B_1\equiv B_{z1}\vec e_z$, $\vec E_1\equiv \vec E_{y1}\vec e_y$) {\em purely compressive} perturbations of the magnetic field, which  describe the fast magneto-acoustic mode with the frequency $\omega^2=k_x^2 \left(c_s^2 + c_a^2\right)$, where $c_s^2=\kappa (T_e+ T_i)/m_i$, and $c_a^2=B_0^2/(\mu_0 m_i n_0)$. This purely compressive nature of the mode within the linear theory is therefore not assumed, but it simply follows from the geometry of the mode.

However, within the nonlinear theory,  the mode does not necessarily remain purely compressive as will be demonstrated below. This important feature has not been explored in the past. It will be shown below that the generalization of the theory is in fact rather straightforward, and it only includes  one additional nonlinear equation for the perpendicular magnetic field which  is, within the given approximations,  identical to the equation for its compressive component.

%\section{Nonlinear  perturbations}

In the nonlinear regime, the term $\left(\nabla\times \vec B\right) \times \vec B$ in the corresponding nonlinear counterpart of Eq.~(\ref{e7}) reads
\[
-\vec e_x\left(B_z\frac{\partial B_z}{\partial x} + B_y\frac{\partial B_y}{\partial x}\right) + \vec e_y B_x\frac{\partial B_y}{\partial x}+\vec e_z B_x\frac{\partial B_z}{\partial x}.\
\]
While writing this expression, we have dropped the subscript $1$ from the perturbed quantities. Clearly, there is an additional perpendicular component $B_y$ which cannot be omitted now. Hence, simply {\em assuming} purely compressive perturbations  in the nonlinear regime may  not always be justified. We shall repeat the derivations by keeping all the terms in the Eqs.~(\ref{ee1})-(\ref{con}), i.e., by retaining the finite electron inertia and collision terms.

Eq.~(\ref{e6}) in the presence of inertia and collision terms now reads
 \be
\vec E=-\frac{\kappa T_e}{e n} \nabla n - \vec v_e\times \vec B - \frac{\nu_{ei} m_e}{e} \left(\vec v_e -\vec v_i\right) - \frac{m_e}{e} \frac{D_e\vec v_e}{Dt}. \label{e66}
\ee
 This is used in Eq.~(\ref{ee1a}) together with the Amp\`{e}re law Eq.~(\ref{ee4}), and,  with the momentum conservation  which in the quasi-neutral system like the present one implies $\nu_{ie}=m_e\nu_{ei}/m_i$. The ion momentum equation now becomes
 \be
 \frac{\partial \vec v_i}{\partial t} + \left(\vec v_i\cdot\nabla\right)\vec v_i=\frac{1}{\mu_0 m_i n}\left(\nabla\times \vec B\right) \times \vec B -
 c_s^2\frac{\nabla n}{n} - \frac{m_e}{m_i} \frac{D_e\vec v_e}{Dt}. \label{e77}
 \ee
The electron velocity in Eq.~(\ref{e77}) is eliminated  by using Eq.~(\ref{ee4})
\[
 \frac{D_e\vec v_e}{Dt}= \left[\ddt + \left(\vec v_i-\frac{\nabla \times \vec B}{\mu_0 e n}\right)\cdot\nabla\right] \left(\vec v_i-\frac{\nabla \times \vec B}{\mu_0 e n}\right)
\]
\be
 \equiv \left(\ddt + \vec v_i\cdot\nabla\right)\left(\vec v_i-\frac{\nabla \times \vec B}{\mu_0 e n}\right)=\frac{m_e}{m_i} \frac{D_i \vec v_i}{Dt}- \frac{m_e}{m_i} \frac{D_i}{Dt}\left( \frac{\nabla \times \vec B}{\mu_0 e n}\right),\label{e77c}
 \ee
 where we have used the fact that $(\nabla\times \vec B)\cdot \nabla=\left[-\vec e_y \partial B_z/(\partial x)+ \vec e_z \partial B_y/(\partial x)\right]\cdot \nabla\equiv 0$.
Eq.~(\ref{e77}) becomes
 \be
 \frac{\partial \vec v_i}{\partial t} + \left(\vec v_i\cdot\nabla\right)\vec v_i=\frac{1}{\mu_0 m_i n}\left(\nabla\times \vec B\right) \times \vec B -
 c_s^2\frac{\nabla n}{n} + \frac{m_e}{m_i} \frac{D_i}{Dt}\left( \frac{\nabla \times \vec B}{\mu_0 e n}\right). \label{e77a}
 \ee
The only approximation used so far  is
\be
1+ m_e/m_i \simeq 1.\label{ap}
\ee
 Applying $\nabla\times$ onto Eq.~(\ref{e66}) and using Eqs.~(\ref{ee3}, \ref{ee4}) yields the following induction equation
\[
\frac{\partial \vec B}{\partial t} = \nabla\times \left[\left(\vec v_i\times \vec B\right) - \frac{1}{\mu_0 e} \left(\frac{\nabla\times \vec B}{n} \times \vec B\right)\right]
\]
 \be
 -\frac{\nu_{ei} m_e}{\mu_0e^2} \nabla \times \left(\frac{\nabla\times \vec B}{n} \right) + \frac{m_e}{e} \nabla\times  \frac{D_e\vec v_e}{Dt}. \label{e88}
 \ee
In the above induction equation, along with the advection, and Hall and Ohm terms,   the correction due to electron inertia has been retained on the right-hand side.

We use Eq.~(\ref{e77c}) to calculate the last term in  Eq.~(\ref{e88}), and this yields
\[
  \frac{\partial \vec B}{\partial t} =-\vec e_y \frac{\partial}{\partial x}\left(v_{ix} B_y- v_{iy} B_x\right)
+\vec e_z \frac{\partial}{\partial x}\left(v_{iz} B_x- v_{ix} B_z\right)
\]
\[
+ \frac{\vec e_y}{\mu_0 e} \frac{\partial}{\partial x}\left(\frac{B_x}{n}\frac{\partial B_z}{\partial x}\right)
- \frac{\vec e_z}{\mu_0 e} \frac{\partial}{\partial x}\left(\frac{B_x}{n}\frac{\partial B_y}{\partial x}\right)
\]
\[
+ \frac{m_e \nu_{ei}}{\mu_0 e^2} \left[ \vec e_y \frac{\partial}{\partial x}\left(\frac{1}{n}\frac{\partial B_y}{\partial x}\right)
+\vec e_z \frac{\partial}{\partial x}\left(\frac{1}{n}\frac{\partial B_z}{\partial x}\right) \right]
\]
\[
+ \frac{m_e}{e} \left\{\ddt \left(-\vec e_y \frac{\partial v_{iz}}{\partial x}+ \vec e_z \frac{\partial v_{iy}}{\partial x}\right)
- \vec e_y \frac{\partial }{\partial x}\left(v_{ix} \frac{\partial v_{iz}}{\partial x}\right)
+\vec e_z \frac{\partial }{\partial x}\left(v_{ix} \frac{\partial v_{iy}}{\partial x}\right)
\right.
\]
\[\left.
+\frac{1}{\mu_0 e}\ddt\left[\vec e_y \frac{\partial}{\partial x} \left(\frac{1}{n} \frac{\partial B_y}{\partial x}\right)
+\vec e_z \frac{\partial}{\partial x} \left(\frac{1}{n} \frac{\partial B_z}{\partial x}\right)\right]
\right.
\]
\be
\left.
+\frac{1}{\mu_0 e} \vec e_y \frac{\partial}{\partial x} \left[v_{ix} \frac{\partial}{\partial x}\left(\frac{1}{n}\frac{\partial B_y}{\partial x}\right)\right]
+ \frac{1}{\mu_0 e} \vec e_z \frac{\partial}{\partial x} \left[v_{ix} \frac{\partial}{\partial x}\left(\frac{1}{n}\frac{\partial B_z}{\partial x}\right)\right]\right\}.
\label{e99}
\ee
From this equation we conclude that
\be
 B_x=0, \label{bx}
 \ee
and thus, the Hall term drops out of the equation. The two remaining components of the magnetic field are consequently described by
\[
\frac{\partial B_y}{\partial t} + \frac{\partial}{\partial x} \left(v_{ix} B_y\right)=
-\frac{m_e}{e}\frac{\partial}{\partial x} \left(\ddt + v_{ix} \frac{\partial}{\partial x}\right) v_{iz}+
\frac{m_e \nu_{ei}}{\mu_0 e^2}\frac{\partial}{\partial x}\left(\frac{1}{n}\frac{\partial B_y}{\partial x}\right)
\vspace{-0.3cm}
\]
\be
+ \frac{m_e}{\mu_0 e^2}\frac{\partial}{\partial x} \left(\ddt + v_{ix} \frac{\partial}{\partial x}\right) \left(\frac{1}{n}\frac{\partial B_y}{\partial x}\right),  \label{ee10}
\ee
\[
\frac{\partial B_z}{\partial t} + \frac{\partial}{\partial x} \left(v_{ix} B_z\right)=
\frac{m_e}{e}\frac{\partial}{\partial x} \left(\ddt + v_{ix} \frac{\partial}{\partial x}\right) v_{iy}+
\frac{m_e \nu_{ei}}{\mu_0 e^2}\frac{\partial}{\partial x}\left(\frac{1}{n}\frac{\partial B_z}{\partial x}\right)
\vspace{-0.3cm}
\]
\be
+ \frac{m_e}{\mu_0 e^2}\frac{\partial}{\partial x} \left(\ddt + v_{ix} \frac{\partial}{\partial x}\right) \left(\frac{1}{n}\frac{\partial B_z}{\partial x}\right). \label{ee10a}
\ee
Within the approximation (\ref{ap}), the first terms on the right-hand sides in Eqs.~(\ref{ee10}, \ref{ee10a}) can be neglected.
This may be seen from the following comparison of terms in (\ref{ee10}). We compare
 the left-hand side [the term $a$] with the first term on the RHS [the term $b$], and the result is
\[
\frac{a}{b}=e B_y/(m_e k v_{iz}).
\]
Here and further we use Eq.~(\ref{7bbb}) (see further in the text) which yields
\[
v_{iz}\simeq \frac{m_e}{m_i} \frac{k B_y}{\mu_0 e n}.
\]
Hence, we have
\[
\frac{a}{b}=\frac{m_i}{m_e}\frac{1}{k^2 \lambda_e^2}\gg 1,  \quad  \lambda_e=c/\omega_{pe}.
\]
Now compare the term $b$ with the last term in (\ref{ee10}):
\[
\frac{b}{c}\simeq \frac{\mu_0 e n v_{iz}}{k B_y}=\frac{m_e}{m_i}.
\]
Finally,  compare $b$ with the second term on the rhs in (\ref{ee10}):
\[
\frac{b}{d}=  \frac{\mu_0 e n \omega v_{iz}}{\nu_{ei} k B_y}=\frac{m_e}{m_i} \frac{\omega}{\nu_{ei}}.
\]
Here $\omega/\nu_{ei}$ is in principle arbitrary but strictly speaking it should be below 1 in order to  justify the use of fluid theory. In view of the mass difference we may conclude that  the first term on the rhs in (\ref{ee10}) is indeed negligible. A similar comparison of terms can be used in Eq.~(\ref{ee10a}), showing that the first term on the rhs in Eq.~(\ref{ee10a}) is negligible as well.

Consequently, what remains are two {\em identical equations} for both the shear and compressional components of the magnetic field $B_y$, $B_z$, obtained within the same approximations:
\[
\frac{\partial B_{\alpha}}{\partial t} + \frac{\partial}{\partial x} \left(v_{ix} B_{\alpha}\right)=
\frac{m_e \nu_{ei}}{\mu_0 e^2}\frac{\partial}{\partial x}\left(\frac{1}{n}\frac{\partial B_{\alpha}}{\partial x}\right)
\vspace{-0.3cm}
\]
\be
+ \frac{m_e}{\mu_0 e^2}\frac{\partial}{\partial x} \left(\ddt + v_{ix} \frac{\partial}{\partial x}\right) \left(\frac{1}{n}\frac{\partial B_{\alpha}}{\partial x}\right), \quad \alpha=y, z. \label{e10c}
\ee
In view of Eq.~(\ref{bx}), from Eq.~(\ref{e77a}) we have the following equations for the ion velocity components
\be
 \left(\ddt + v_{ix}\frac{\partial}{\partial x}\right)v_{ix} = -c_s^2 \frac{1}{n} \frac{\partial n}{\partial x} -\frac{1}{2\mu_0 m_i n}\frac{\partial}{\partial x}\left(B_y^2+ B_z^2\right), \label{7b}
 \ee
\be
 \left(\ddt + v_{ix}\frac{\partial}{\partial x}\right)v_{iy} = -\frac{1}{\mu_0 e} \frac{m_e}{m_i}\left(\ddt + v_{ix}\frac{\partial}{\partial x}\right) \left(\frac{1}{n}\frac{\partial B_z}{\partial x}\right), \label{7bb}
 \ee
 \be
 \left(\ddt + v_{ix}\frac{\partial}{\partial x}\right)v_{iz} = \frac{1}{\mu_0 e} \frac{m_e}{m_i}\left(\ddt + v_{ix}\frac{\partial}{\partial x}\right) \left(\frac{1}{n}\frac{\partial B_y}{\partial x}\right).\label{7bbb}
 \ee
The ion continuity rewritten here reads
\be
 \frac{\partial n}{\partial t} + \frac{\partial}{\partial x}\left(n v_{ix}\right)=0.\label{con2}
 \ee
Within the approximation (\ref{ap}), the equations (\ref{7b}), (\ref{con2})  together with the {\em two equations} (\ref{e10c}) make a complete closed set.
This set of equations  contains both  the compressional and  the perpendicular components of the magnetic field $B_z$, $B_y$.

Observe that Eq.~(\ref{e10c}) is formally correctly satisfied with the trivial case $B_y\equiv 0$,  which then reduces the equations to those studied so far in Refs.~\cite{aa}-\cite{shi}. However, the magnetic field equation is identical for both $B_y$ and $B_z$, and thus the same  can be done with the $B_z\equiv 0$, and what remains is a closed set which determines the nonlinear perpendicular (shear) component of the magnetic field $B_y$, with a plethora of possible solutions that have not been explored in the past. This is even more so if the nontrivial case is studied for both components. Clearly, neglecting one or the other component without justification may lead to unphysical singular solution.

In some cases, solving the complete set of equations (\ref{e10c}), (\ref{7b}), (\ref{con2})  can be  rather straightforward, and it can easily be reduced
to the procedure used for the purely compressive perturbations. This is most clearly seen in the case of cold plasma and in the limit of massless electrons.
From Eqs.~(\ref{e10c}), (\ref{con2})  we obtain the frozen-in condition for both components $B_y, B_z$ separately:
\be
B_z=a_1 n, \label{f1}
\ee
\be
B_y=a_2 n, \label{f2}
\ee
where $a_1, a_2$ are arbitrary constants. This further implies that $B_y=a_3 B_z$; using this in Eq.~(\ref{7b})
we have
\be
\frac{\partial v_{ix}}{\partial t} + v_{ix} \frac{\partial v_{ix}}{\partial x} + c_1 \frac{\partial B_z}{\partial x}  =0, \quad
c_1= \frac{a_1(a_3^2 + 1)}{\mu_0 m_i}. \label{f3}
\ee
This equation is coupled with
\be
\frac{\partial B_z}{\partial t} + B_z \frac{\partial v_{ix}}{\partial x} + v_{ix} \frac{\partial B_z}{\partial x}=0.
\label{f4}
\ee
This set of equations can be solved exactly following the procedure suggested by Stenflo {\em et al.}  \cite{s1} for the purely compressive perturbations.
We may take
\be
B_z=\left(b_1 + \frac{b_2 v_{ix}}{2}\right)^2, \label{f5}
\ee
where $b_1, b_2$ are some arbitrary constants. Using this in Eq.~(\ref{f4}), it turns out that Eq.~(\ref{f4}) becomes identical to Eq.~(\ref{f3}) provided that
\be
c_1\equiv \frac{1}{b_2^2}. \label{f6}
\ee
The resulting equation for $v_{ix}$ reads
\be
\frac{\partial v_{ix}}{\partial t} + \frac{3}{2} v_{ix} \frac{v_{ix}}{\partial x} + \frac{b_1}{b_2} \frac{v_{ix}}{\partial x}=0.
\label{f7}
\ee
This equation is identical to Eq.~(4) from Ref. \cite{s1} obtained for compressive perturbations. It describes steepening of the wave profile and shock wave formation in cold plasma. Having the solution for the speed $v_{ix}$, we can further easily find the magnetic field profile by calculating its components with the help of Eqs.~(\ref{f1}, \ref{f2}, \ref{f5}). The solution is therefore physically much different compared to Ref.~\cite{s1} because  it  implies
both compressional and shear components.
A similar procedure  can be done  for soliton description within the reductive perturbation technique  used in many references, see the latest in Refs.~\cite{hu3, shi}.

\section{Conclusion}

It may be concluded that the purely  compressive magneto-acoustic perturbations studied in Refs.~\cite{aa}-\cite{shi} and in many others are indeed mathematically possible, yet they are physically very specific and  obtained after assuming the trivial value for the perpendicular perturbation. The later is  in fact described by a completely identical nonlinear equation  derived for the first time in the present work.  In general case the compressive and perpendicular perturbations are coupled, the corresponding nonlinear equations contain no small parameters that would support the fact that the perpendicular part is always explicitly omitted. Therefore, in real situations the purely compressive solutions may be far less abundant than expected. After half a century of investigations of the purely compressive mode, a more complete theory is needed which would include both perturbed components of the magnetic field.  Such a generalized theory is described here. It contains  a rather straightforward derivation of a complete set of equations, which includes an  additional equation for the shear component of the magnetic field. Surprisingly, this equation turns out to be exactly the same as the usual equation for the compressive part of the magnetic field, and it is obtained using the same approximations. This implies that the previously used standard procedures of solving nonlinear equations for the compressive nonlinear magnetic structures can, at least in some cases, be used also in the presented generalized set of equations. One example of that kind is described in the present work.

\end{document}